\documentclass[12pt, letterpaper]{article}
\newcommand{\lsim}{\raisebox{-0.13cm}{~\shortstack{$<$ \\[-0.07cm]
      $\sim$}}~}
\newcommand{\gsim}{\raisebox{-0.13cm}{~\shortstack{$>$ \\[-0.07cm]
      $\sim$}}~}
\usepackage[top=1in, bottom=1.0in, left=1in, right=1in]{geometry}
\usepackage[utf8]{inputenc}
\usepackage{booktabs}
\usepackage{hyperref}
\usepackage{graphicx}

\title{The Plane's The Thing: The Case for Wide-Fast-Deep Coverage of the Galactic Plane and Bulge}
\date{November 28, 2018}

\begin{document}

\maketitle

\begin{abstract}
We argue that the exclusion of the Galactic Plane and Bulge from the uniform wide-fast-deep (WFD) cadence is fundamentally inconsistent with two of the main science drivers of LSST: 
Mapping the Milky Way and Exploring the Transient Optical Sky. We outline the philosophical basis for this claim and then describe a number of important science goals that can only be addressed by WFD-like coverage of the Plane and Bulge.
\end{abstract}

\section{White Paper Information}
Jay Strader (Michigan State; {\tt strader@pa.msu.edu}),

\noindent
Elias Aydi (Michigan State),
Christopher Britt (STScI),
Adam Burgasser (UCSD),
Laura Chomiuk (Michigan State),
Will Clarkson (UM-Dearborn),
Brian D. Fields (Illinois),
Poshak Gandhi (Southampton),
Leo Girardi (Padova),
John Gizis (Delaware), 
Jacob Hogan (Illinois),
Michael A.~C.~Johnson (Southampton),
James Lauroesch (Louisville),
Michael Liu (Hawaii), 
Tom Maccarone (Texas Tech),
Peregrine McGehee (College of the Canyons),
Dante Minniti (Universidad Andr\'{e}s Bello),
Koji Mukai (NASA/Goddard),
Alexandre Roman-Lopez (La Serena),
Simone Scaringi (Texas Tech),
Jennifer Sobeck (Washington),
Kirill Sokolovsky (Michigan State),
C. Tanner Murphey (Illinois),
Xilu Wang (Notre Dame)

\vspace{2mm}
\noindent
on behalf of the Stars, Milky Way, and Local Volume Collaboration

\begin{enumerate} 
\item {\bf Science Category:} 

Exploring the Transient Optical Sky

Mapping the Milky Way

\item {\bf Survey Type Category:} Wide-fast-deep
\item {\bf Observing Strategy Category:} 
An integrated program with science that hinges on the combination of pointing and detailed 
observing strategy
\end{enumerate}

\clearpage

\section{Scientific Motivation}

\vspace{-0.15cm}
The primary argument of this white paper is \emph{philosophical}: that exclusion of the Galactic Plane/Bulge from the uniform cadence is inconsistent with two of the main science drivers of LSST: Mapping the Milky Way and Exploring the Transient Optical Sky. We describe this claim in broad terms and then describe a number of science goals that can only be addressed by wide-fast-deep (WFD)-like coverage of the Galactic Plane and Bulge. The science goals listed here are necessarily incomplete, and other white papers make complementary points.

\vspace{-0.25cm}
\subsection{The Plane/Bulge in the Baseline Cadence}

\vspace{-0.1cm}
In the baseline cadence, the relevant region excluded from the WFD coverage is defined $10^{\circ}$ above and below the Galactic Center, with two lines extending from each point to $b=0^{\circ}$ at $l=90^{\circ}$ and $l=270^{\circ}$ (see visualization of this area in Figure 1). Instead of the $\sim 825$ visits for an average region with WFD coverage, this region toward the inner Galaxy would only receive 30 visits total per filter over the 10-year survey, or 3 visits per year per filter on average, as part of a special proposal. Within the nominal declination range of the WFD part of the baseline cadence, this area represents about 8\% of the WFD sky coverage, but \emph{over half of the stellar mass of the Galaxy}. Obviously, for populations concentrated in the Plane, such as young stars, X-ray binaries, and supernovae, the fraction excluded is much higher and approaches unity.

While crowding (confusion) will limit the co-added depth in some filters at low Galactic latitude, this is irrelevant for most of the compelling Galactic science cases, which focus on variables, transients, and other epoch-level science such as proper motions and parallax.

\vspace{-0.25cm}
\subsection{Black Holes and Neutron Stars}

\vspace{-0.1cm}
Approximately 200 low-mass X-ray binaries (neutron stars or black holes with low-mass companions) are known in the Galaxy (Liu et al.~2007), but the total population is likely many thousands. Many questions central to modern astrophysics can only be answered by enlarging this sample: which stars produce neutron stars and which black holes; whether there is a true mass gap between neutron stars and black holes; if supernova explosions give large black hole kicks; the origins of observed gravitational wave sources.

Nearly all of these X-ray binaries were discovered when they went into outburst. Since outburst recurrence times likely extend to 10$^2$--10$^4$ yr (e.g., Piro \& Bildsten 2002), most binaries will not show such outbursts over human timescales. LSST imaging offers a different route to discovery with fewer selection effects.
In quiescence, low-mass X-ray binaries can be identified as optical variables that show  double-humped ellipsoidal variations of typical amplitude $\sim 0.2$ mag due to the tidal deformation of the secondary. Photometry alone gives constraints on the mass ratio and inclination, and together with spectroscopy from 4 to 10-m class telescopes, one can determine the mass of the black hole or neutron star primary.

Johnson et al.~(2018) show that the extension of WFD to the Plane/Bulge allows the characterization of orbital periods for essentially all quiescent X-ray binaries down to $r\sim23$, but that the baseline cadence will recover $\lsim 25$\% of these periods (Figure 2). Hence extending WFD to the Plane/Bulge would \emph{quadruple} the number of quiescent black hole binaries detected. This is consistent with the observation that of the known Galactic black holes/candidates, 73\% are in the Galactic region excluded from WFD (Tetarenko et al.~2016).

\vspace{-0.25cm}
\subsection{Novae} 

\vspace{-0.1cm}
Classical novae, thermonuclear runaways triggered in the hot material accreted onto a white dwarf from a companion, are the most common distinct explosions in the Galaxy, with an observed rate of $\sim 10$--15 per year (e.g., Shafter 2017).

A central ongoing question in astrophysics is whether accreting white dwarfs are progenitors for Type Ia SNe. For this to be true, white dwarfs must \emph{grow} in mass as they accrete, rather than stagnating or losing mass through novae. This question can be answered by measuring the orbital period of the binary before and after a nova, which can be used to quantify the ejected mass. The fractional period change expected is in the range $10^{-4}$ to $10^{-6}$ (Patterson et al.~2017), which is measurable with a WFD-like cadence but not with the limited baseline cadence (see Figure 3).

Except for the tiny number of recurrent novae, no current or planned surveys can provide pre-explosion orbital periods for typical novae, which have faint red dwarf donors and $r \gsim 18$. WFD-like cadence would allow precise pre-explosion orbital periods to be determined for many novae. Since 91\% of novae since 2008 have occurred in the currently excluded Plane/Bulge region, this project is only possible with WFD-like cadence of this region.

\vspace{-0.3cm}
\subsection{A Galactic Supernova}

\vspace{-0.15cm}
Galactic supernovae (SNe) are rare, but a SN in the Milky Way would be among the most important astronomical events of our lifetime. Massive star deaths lead to a core-collapse rate in the Galaxy estimated to be $3.2^{+7.3}_{-2.6}$ ($2\sigma$) per century. The rate of Type Ia explosions is $1.4^{+1.4}_{-0.8}$ per century (Adams et al.~2013). Thus, within a 10 year LSST survey, the Poissonian probability of {\em any} Galactic SN explosion is $\sim 37\%$ ($\sim 17\%$ to $70\%$
at the $1\sigma$ level).

{\bf Core-collapse SNe:} Under most scenarios, it is unlikely that LSST would discover the SN. The real value of LSST would be pre-SN observations. These would be critical to identifying the progenitor, recording precursor outbursts, and probing its possible binarity.

{\bf Type Ia SNe:} In sharp contrast to the spectacular multi-messenger outburst that would accompany a core-collapse SN, a Galactic Type Ia supernova may go initially unnoticed. For distances $> 1 \ \rm kpc$ it will probably be missed in neutrinos (Odrzywolek \& Plewa 2011), and for a double degenerate progenitor there may be no bright X-ray/$\gamma$-ray shock breakout signature (Nakar \& Sari 2012). Since SNe Ia with short delay times may well occur in highly-extinguished regions (Adams et al.~2013), deep frequent imaging with LSST could discover or allow the characterization of such an SN. 

LSST is the only facility with the ability to identify or characterize nearly any Galactic SN, but only if a  WFD-like cadence is used in the Plane/Bulge region (Figure 1).

\vspace{0.25cm}
\noindent
{\bf \large Other Science:} {\bf New Dwarf Galaxies}: Among the most promising ways to discover new dwarf galaxies is via RR Lyrae variables, which trace old metal-poor populations. The area behind the Plane/Bulge of the Galaxy is basically empty of known dwarfs---is this real (with important implications for galaxy formation) or simply a selection effect? The Nov 2018 discovery of Antlia 2 (now the largest faint Galactic dwarf; Torrealba et al.~2018) at low Galactic latitude shows the pressing discovery space available in this area.


\vspace{1.2mm}
\noindent
{\bf Ultracool stars}: WFD-like cadence is necessary to find complete samples of ultracool dwarfs via color, proper motion, and parallax. Dense regions enable AO-based follow-up. \newpage

\begin{figure}
\begin{center}
\vspace{-0.5cm}
\hspace*{1.1cm}\includegraphics[width=6.4in]{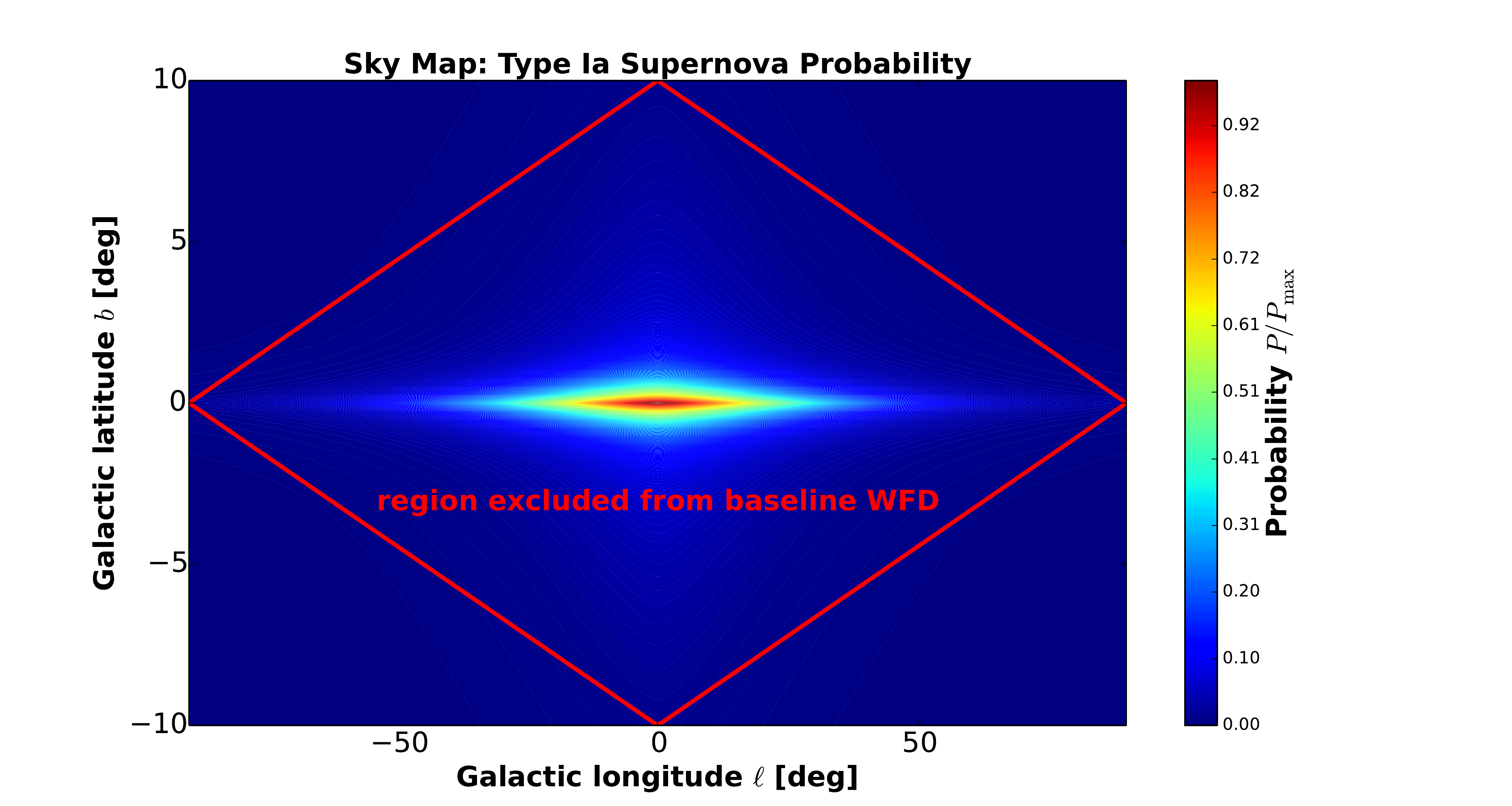}
\caption{\small This figure shows the stellar density of the Milky Way in Galactic coordinates, with the colors representing the relative odds of the occurrence
of a Galactic Type Ia SN (red=high, blue=low). Nearly all ($> 90$\%) of the probability is contained within the red diamond, representing the Plane/Bulge region that is
excluded from the WFD survey under the current baseline cadence.}
\label{Fig:plot2}
\end{center}
\end{figure}

\begin{figure}[h!]
\vspace{-2.6cm}
\begin{center}
  \includegraphics[width=4.6in,angle=270]{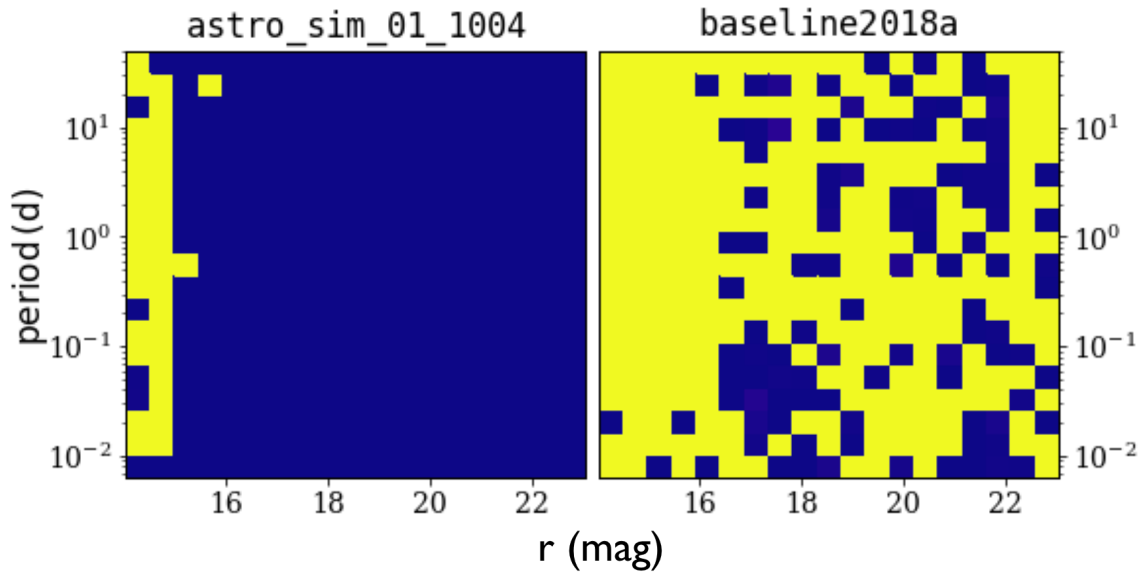}
\vspace{-2.2cm}
\caption{\small The recovery of low-mass X-ray binary orbital periods as a function of mean $r$ mag for realistic light curves for a Galactic field using two LSST cadences: the baseline cadence ({\tt baseline2018a}) with limited coverage of the Plane/Bulge, and a sample cadence in which that region has WFD cadence ({\tt astro\_sim\_01\_1004}). The color scale shows the significance of recovery (blue=high, yellow=low). It is clear that with a WFD-like cadence in the Plane/Bulge, orbital periods can be determined for nearly all well-behaved quiescent black hole and neutron star X-ray binaries down to $r\sim23$, while in the baseline cadence, most would have an insufficient number of visits to determine orbital periods. Figure adapted from Johnson et al.~(2018).}
\label{Fig:plot1}
\end{center}
\vspace{-2.5cm}
\end{figure}

\newpage

\vspace{.6in}

\section{Technical Description}

\subsection{High-level description}

Add the currently excluded Plane/Bulge region to the uniform (wide-fast-deep) cadence.


\subsection{Footprint -- pointings, regions and/or constraints}

The excluded region to be added to WFD is defined as $10^{\circ}$ above and below the Galactic Center, with two lines from each point extending to $b=0^{\circ}$ at $l=90^{\circ}$ and $l=270^{\circ}$.

\subsection{Image quality}

It would be reasonable to de-prioritize the densest fields in $> 1"$ seeing since this aggravates confusion issues. Else  
the image quality constraints should be the same as WFD.

\subsection{Individual image depth and/or sky brightness}

No extra constraints beyond the standard lunar avoidance for WFD.

\subsection{Co-added image depth and/or total number of visits}

Uniform WFD cadence.

\subsection{Number of visits within a night}

No constraints.

\subsection{Distribution of visits over time}

No strong constraints; something between a uniform and a ``soft" rolling cadence will be reasonable. See technical description/trade-offs below.

\subsection{Filter choice}

Standard WFD filters.

\subsection{Exposure constraints}

Standard WFD exposure. The efficiency gains of a single 30-sec exposure over 2 15-sec exposures are large and hence the overall LSST pie grows substantially ($\sim 6$--7\%).
Adopting the longer exposure time obviously extends the saturation limit to brighter magnitudes, but in general similar data already exist from other all-sky surveys such as ASAS-SN. Hence we strongly advocate for single 30-sec exposures.

\subsection{Other constraints}

No constraints.

\subsection{Estimated time requirement}

In previous simulations of this proposal, the area of WFD is increased by $\sim 8$\% but the total number of WFD visits only increased by $\sim 1.6$--2\% since the priority of WFD as a whole is competed against other proposals. This has led to co-added depth over all of WFD that is lower by 0.01--0.03 mag depending on the filter. Hence it is reasonable to take 1.6--2\% of the baseline WFD survey as an estimate of the time requirement.

\vspace{.3in}

\begin{table}[ht]
    \centering
    \begin{tabular}{l|l|l|l}
        \toprule
        Properties & Importance \hspace{.3in} \\
        \midrule
        Image quality &   2  \\
        Sky brightness & 2 \\
        Individual image depth &  2 \\
        Co-added image depth & 3  \\
        Number of exposures in a visit   &  3 \\
        Number of visits (in a night)  &  3  \\ 
        Total number of visits &  1 \\
        Time between visits (in a night) & 3 \\
        Time between visits (between nights)  & 1.5  \\
        Long-term gaps between visits & 2 \\
        Other (please add other constraints as needed) & \\
        \bottomrule
    \end{tabular}
    \caption{{\bf Constraint Rankings:} Summary of the relative importance of various survey strategy constraints. Please rank the importance of each of these considerations, from 1=very important, 2=somewhat important, 3=not important.} 
        \label{tab:obs_constraints}
\end{table}

\subsection{Technical trades}

Given that the primary science goals for the Plane/Bulge are epoch-level and not co-added (though there certainly is useful co-added science; see the Gonzalez et al white paper), the most important parameter is the total number of visits and not the co-added depth.

Considering the distribution of visits: for some science cases a more uniform coverage is desirable (e.g., discovery of new transients such as a Galactic supernova or pre-explosion nova periods), while for other science cases (determining the period of short-period binaries) some flavor of a rolling cadence would give similar results to a uniform cadence. A hybrid approach may be the best compromise with other science goals. A ``hard" rolling cadence where fields receive most of their visits in a short period of time are likely to fare poorly by some metrics, such as determining the periods of binaries with $\gsim 1$ year periods, or Galactic supernova discovery.

We note that in the current limited Plane proposal, there is no requirement for 2 visits to be taken per night to identify fast moving objects. Given the increasing interest (relevant to ``Planet 9") in finding Solar System objects outside of the ecliptic, it seems reasonable to adopt the standard WFD strategy of 2 visits per night in the Plane/Bulge as well.

Considering the footprint: here we have argued for the inclusion of the Plane/Bulge within the nominal WFD declination limits, while other proposals will argue to expand WFD to higher declinations. Given that (a) comparatively little of the stellar mass of the Galaxy is located in the Plane at northern declinations, (b) the highest declination fields would necessarily be observed at high airmass, and (c) other surveys are likely to be operating in the northern hemisphere, we do not advocate for a Pan-STARRS-like cadence to high declination. A marginal change in the declination limit of the WFD survey, or the retention of the special Plane proposal solely at high declination, might well be reasonable.

Considering filters: while some science cases might marginally benefit from a restricted subset of filters over the Plane/Bulge region, our view is that this is outweighed by the science cases that do use the bluer filters, such as  pre-explosion variability of a Galactic supernova. While $u$ and $g$ are obviously affected by extinction very close to the Plane, paradoxically they will also aid in the construction of a precise 3-D reddening map over the entire Galaxy (including the Plane/Bulge), benefitting all Plane/Bulge science, as has been done for Pan-STARRS1 (Green et al.~2018). Given a variety of arguments for or against minor changes to the filter set in the Plane/Bulge, it is reasonable that the survey principle of the advantage of a homogenous dataset should prevail.

\section{Performance Evaluation}

It is worth emphasizing that while individual metrics can assess the effectiveness of different cadences to achieve different science goals, they cannot \emph{weigh} science cases against each other. In a very real sense, a metric for a constraint on the dark energy equation of state and a metric on the number of stellar-mass black holes discovered by LSST are incommensurable.\\

The main heuristics are the WFD-related ones relevant for individual visits: the total number of visits per filter, the median and maximum inter-night visit per filter, and the general ones involving image quality, lunar avoidance.

The specific science cases in this white paper, which are a small subset of the science enabled by WFD-like coverage the Plane/Bulge, have detailed figures of merit, some of which have already been evaluated against potential cadences. For low-mass X-ray binaries, the scientific metric is the number of binaries with orbital periods recovered; the number of new black holes or neutron stars discovered will scale directly with this metric. As discussed above, Johnson et al.~(2018) assess this metric for different potential cadences and find that within the Plane/Bulge (where about 73\% of known black holes are located), with the baseline cadence few or no binaries will have correctly recovered periods. However, a WFD-like cadence in the Plane/Bulge will enable precise orbital periods to be determined for nearly all well-behaved quiescent black hole and neutron star X-ray binaries down to $r\sim23$ (Figure~1). The nova science case has a similar metric, which is the number of novae for which post-nova fractional period changes of order $10^{-5}$ are measurable. We have evaluated this metric for the previous baseline cadence ({\tt minion\_1016}) and for one with WFD-like coverage of the Plane/Bulge ({\tt astro\_sim\_01\_1004}), finding that this precision is measurable for most observable systems in the latter case and nearly none in the baseline cadence (Figure 3).

Several figures of merit have been written for the Galactic supernova science cases: for the presence of pre-explosion variability, there is FoM 3.1 in Chapter 4 of
``Science-Driven Optimization of the LSST Observing Strategy" (Marshall et al.~2017), and for Type Ia supernova discovery the metric will depend inversely
on the median and maximum inter-night gaps.

\begin{figure}
\begin{center}
\vspace{-1.0cm}
  \includegraphics[width=4in]{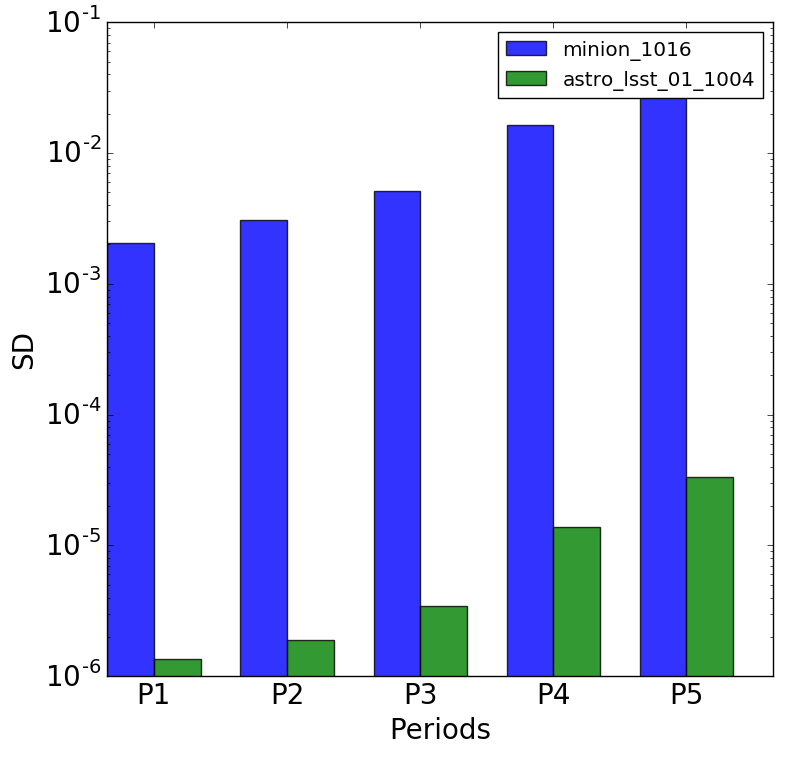}
\caption{\small This histogram shows the typical fractional accuracy of periods for two cadences: the previous baseline cadence ({\tt minion\_1016}) and one with WFD-like coverage of the Plane/Bulge ({\tt astro\_sim\_01\_1004}) ``P1" to ``P5" represent monotonically increasing sample periods: 0.15 d, 0.22 d, 0.40 d, 1.54 d, and 3.30 d. It is clear that the accuracy necessary to measure the expected period change post-nova is possible in a WFD-like cadence but not in the baseline cadence.}
\label{Fig:plot3}
\end{center}
\end{figure}


\section{Special Data Processing}

The current special proposal for the Plane/Bulge already includes these fields and hence the challenges of high-density regions with many stars already need to be addressed. Hence this proposal should not represent additional effort required by Data Management.

\section{References}

{\bf Adams, S.}, et al. 2013, ApJ, 778, 164\\
{\bf Green, G.}, et al. 2018, MNRAS, 478, 651\\
{\bf Johnson, M.}, et al. 2018, MNRAS, in press (arXiv:1809.09141)\\
{\bf Liu, Q.}, et al. 2007, A\&A, 469, 807\\
{\bf Marshall, P.}, et al., 2017, (arXiv:1708.04058)\\
{\bf Nakar, E.} \& Sari, R. 2012, ApJ, 747, 88\\
{\bf Odrzywolek, A.} \& Plewa, T. 2011, A\&A, 529, 156\\
{\bf Patterson J.}, et al. 2017, MNRAS, 466, 581\\
{\bf Piro A.} \& Bildsten, L.\ 2002, ApJL, 571, 103\\
{\bf Shafter, A.} 2017, ApJ, 834, 196\\
{\bf Tetarenko, B.}, et al. 2016, ApJS, 222, 15\\
{\bf Torrealba, G.}, et al. 2018, MNRAS, submitted (arXiv:1811.04082)

\end{document}